\documentclass[aps,pra,showpacs,notitlepage,floatfix,
twocolumn,superscriptaddress]{revtex4-1}
\usepackage{bbm}
\usepackage{mathrsfs}
\usepackage{epsfig}

\usepackage{soul,xcolor}
\usepackage{graphicx}
\graphicspath{{./figures/}}

\usepackage{amsfonts}
\usepackage{amsthm}
\usepackage[figuresright]{rotating}
\usepackage{amssymb}
\usepackage{amsmath}
\usepackage{dcolumn}
\usepackage{float}
\usepackage{bm}
\usepackage{braket}
\usepackage[normalem]{ulem}

\usepackage[colorlinks,linkcolor=blue,anchorcolor=blue,citecolor=blue,urlcolor=blue]{hyperref}

\usepackage{physics}

\newcommand{\BPS}{\mbox{\tiny BPS}}
\newcommand{\PS}{\mbox{\tiny PS}}
\newcommand{\MPS}{\mbox{\tiny MPS}}
\newcommand{\generic}{\mbox{\tiny gen.}}

\def\be{\begin{equation}} \def\ee{\end{equation}}
\def\bea{\begin{equation}\begin{aligned}} \def\eea{\end{aligned}\end{equation}}

\begin{document}

\title{Entanglement and tensor networks for supervised image classification}
\author{John Martyn}\affiliation{X, The Moonshot Factory, Mountain View, CA 94043, USA}
\author{Guifre Vidal}\affiliation{X, The Moonshot Factory, Mountain View, CA 94043, USA}
\author{Chase Roberts}\affiliation{X, The Moonshot Factory, Mountain View, CA 94043, USA}
\author{Stefan Leichenauer}\affiliation{X, The Moonshot Factory, Mountain View, CA 94043, USA}

\begin{abstract}
Tensor networks, originally designed to address computational problems in quantum many-body physics, have recently been applied to machine learning tasks. However, compared to quantum physics, where the reasons for the success of tensor network approaches over the last 30 years is well understood, very little is yet known about why these techniques work for machine learning. The goal of this paper is to investigate entanglement properties of tensor network models in a current machine learning application, in order to uncover general principles that may guide future developments. We revisit the use of tensor networks for supervised image classification using the MNIST data set of handwritten digits, as pioneered by Stoudenmire and Schwab [Adv. in Neur. Inform. Proc. Sys. 29, 4799 (2016)]. Firstly we hypothesize about which state the tensor network might be learning during training. For that purpose, we propose a plausible candidate state $\ket{\Sigma_{\ell}}$ (built as a superposition of product states corresponding to images in the training set) and investigate its entanglement properties. We conclude that $\ket{\Sigma_{\ell}}$ is so robustly entangled that it cannot be approximated by the tensor network used in that work, which must therefore be representing a very different state. Secondly, we use tensor networks with a \textit{block product} structure, in which entanglement is restricted within small blocks of $n \times n$ pixels/qubits. We find that these states are extremely expressive (e.g. training accuracy of $99.97 \%$ already for $n=2$), suggesting that long-range entanglement may not be essential for image classification. However, in our current implementation, optimization leads to over-fitting, resulting in test accuracies that are not competitive with other current approaches.
\end{abstract}

\maketitle

\section{Introduction}

Over the past decade, research in artificial intelligence has unveiled a symbiotic relationship between physics and machine learning. For instance, neural networks have been used to locate phase transitions in spin models and even develop equations of motion from empirical data \cite{2016PhRvB..94s5105W, 2018arXiv181010525W, 2020arXiv200309905K}. On the flip side, tensor networks, initially devised to model quantum many-body states, have been successfully applied to supervised learning tasks, such as the recognition of handwritten digits, medical image classification, and anomaly detection  \cite{2016arXiv160505775M, 2018arXiv180605964G, 2018arXiv180100315S, 2020arXiv200410076S, 2020arXiv200413747T, 2019arXiv190606329E, 2020arXiv200602516W, 2019PhRvB..99o5131C, 2020arXiv200108286R}.

Inspired by the well-documented success of tensor networks in quantum many-body physics over the last 30 years, these machine learning studies \cite{2016arXiv160505775M, 2018arXiv180605964G, 2018arXiv180100315S, 2020arXiv200410076S, 2020arXiv200413747T, 2019arXiv190606329E, 2020arXiv200602516W, 2019PhRvB..99o5131C, 2020arXiv200108286R} have incorporated networks such as the matrix product state (MPS) \cite{2006quant.ph..8197P, PhysRevLett.69.2863, fannes1992finitely, PhysRevB.55.2164, 2003PhRvL..91n7902V, 2004PhRvL..93d0502V}, the tree tensor network \cite{2006PhRvA..74b2320S, PhysRevB.82.205105}, and multiscale entanglement renormalization ansatz \cite{2008PhRvL.101k0501V, 2009PhRvB..79n4108E}. Introductions to tensor networks in the language of machine learning can be found in Refs. \cite{2014arXiv1407.3124C, oseledets2011tensor}. 
It is important to keep in mind that tensor network models are \textit{linear} models with an input space that is exponentially large in the number of features (for instance, the number of pixels in an image). The data is first embedded (non-linearly!) in this exponentially large vector space (see Sec. \ref{sec:II} for a discussion of this embedding). Thanks to the embedding, linear models in this vector space have strong expressive power. However, they depend on exponentially many parameters --that is, they are afflicted by the \textit{curse of dimensionality}. The magic of tensor networks is that they offer a manageable, efficient description of a restricted class of linear models in this high-dimensional vector space. Linear models restricted to be of the tensor network class appear to still retain a significant amount of their expressive power.

One might thus expect tensor networks to work well in machine learning due to their expressive power and the observation that patterns in real-world data are relatively simple \cite{2017JSP...168.1223L}. In current studies \cite{2016arXiv160505775M, 2018arXiv180605964G, 2018arXiv180100315S, 2020arXiv200410076S, 2020arXiv200413747T, 2019arXiv190606329E, 2020arXiv200602516W, 2019PhRvB..99o5131C, 2020arXiv200108286R}, a tensor network architecture is selected, and its tensors are optimized so as to minimize a loss function on a training set. Subsequently, its performance is evaluated on the test sets. These methods have been shown to work surprisingly well; for instance, the MPS model can achieve test accuracies upwards of $99\%$ on the MNIST data set of handwritten digits \cite{2016arXiv160505775M}.

In quantum physics, the success of tensor networks such as MPS is ultimately based on a well understood fact. Namely, tensor networks share an important structural property with the quantum states (e.g. ground states of local Hamiltonians) that they try to approximate. This property is known as the \textit{area law} of entanglement, \cite{2017JPhA...50v3001B, 2014AnPhy.349..117O}. How about in machine learning? Suppose we use the above embedding into an exponentially large vector space, so as to encode the data into a quantum state (see Sec. \ref{sec:II} for a definition of quantum states). What property do typical data sets have that, upon this embedding into a quantum state, might play an analogous role to that of the area law in quantum physics? Although a direct answer seems elusive, it must have to do with correlations, e.g., between neighboring pixels in an image. After embedding a set of images in an exponentially large vector space, these correlations are formally related to entanglement in quantum physics. The goal of this paper is to explore the entanglement properties of tensor networks when used for machine learning. For concreteness, we focus on supervised image classification of the MNIST dataset of handwritten digits, following Ref. \cite{2016arXiv160505775M}, and present two main results. 

The first result refers to the amount of entanglement in tensor networks for machine learning. We consider an embedding of the MNIST images, which are comprised of $28 \times 28$ pixels, in a state of a square lattice of $28 \times 28$ qubits. We then introduce a \textit{sum} state, $\ket{\Sigma_{\ell}}$, of the $28 \times 28$ qubits, built as a linear combination of embedded images. (Here $\ell$ is a class label that will be described later on). We initially regarded the sum state $\ket{\Sigma_{\ell}}$  as a plausible candidate for what the MPS model in Ref. \cite{2016arXiv160505775M} might be attempting to learn. We found, however, that the sum state $\ket{\Sigma_{\ell}}$ has very large amounts of entanglement, making it impossible for the MPS model to learn it, even approximately. We thus conclude that the MPS successfully used in Ref. \cite{2016arXiv160505775M} for image classification must represent some very different, less entangled state of the $28 \times 28$ qubits.

The above result referred to the \textit{amount} of entanglement in a particular state. Our second result refers instead to the \textit{range}, in space, of entanglement. Entanglement correlates different parts of the system, and we may ask about how distant these parts are. For this purpose, we divide the 28 $\times$ 28 qubits pixels into blocks, indexed by $b$, of $n\times n$ adjacent qubits, and consider tensor networks that represent a state  $\ket{\Psi^{\BPS}_{\ell}} = \bigotimes_{b} |\psi_{\ell}^{b}\rangle$ that factorizes as the product of individual states $\ket{\psi_{\ell}^{b}}$ for each of the blocks. By construction, this \textit{block product state} (BPS) wavefunction $\ket{\Psi^{\BPS}_{\ell}}$ only has entanglement within each block $b$. That is, $\ket{\Psi^{\BPS}_{\ell}}$ only has \textit{short range entanglement}. Our second result is the realization that this simple tensor network with only short range entanglement within each block is already extremely expressive, in that it leads to very high accuracy when classifying the training set even for small blocks made of $2 \times 2$ qubits. However, the optimization of the model results in significant over-fitting. Indeed, the trained model generalizes poorly to the test set, for which the accuracy is not yet competitive. We are still hopeful that by training the model with a different optimization algorithm, we may obtain much better test accuracies, although we leave this for subsequent explorations.
 
The rest of the paper is organized as follows. In Sec. \ref{sec:II}, we summarize the general set-up (embedding, tensor network, loss function, etc) used in previous studies, and then describe our own set-up, which differs slightly from those of previous studies. In Sec. \ref{sec:III} we introduce the sum state $\ket{\Sigma_{\ell}}$ and study its entanglement properties, to conclude that it is too entangled to be learned by the MPS used in Ref. \cite{2016arXiv160505775M}. In Sec. \ref{sec:IV} we introduce the block product state $\ket{\Psi_{\ell}^{\BPS}}$, which we realize in not one but two different tensor network models (dubbed \textit{nearest neighbor} BPS, and \textit{snake} BPS) and analyze how the two different realizations perform. Finally, in Sec. \ref{sec:V} we summarize our results. 

\section{Protocol for supervised image classification with tensor networks} \label{sec:II}

In this section we discuss the methodology of applying tensor networks to supervised learning, focusing on the problem of image classification. We first summarize the approach laid out in Ref. \cite{2016arXiv160505775M}, after which we discuss our modified protocol. 

\subsection{Previous Work} \label{sec:PrevWork}

Previous works \cite{2016arXiv160505775M, 2020arXiv200413747T, 2020arXiv200410076S, 2018arXiv180100315S, 2019arXiv190606329E, 2020arXiv200602516W} that perform supervised learning with tensor networks employ the following protocol. For concreteness, consider supervised learning of scale-gray images, where each image is made of $N$ pixels. For instance, in the MNIST data set of handwritten digits, each image is made of $N=28\times 28 = 784$ pixels. The data of an image is stored in a vector $x \in V$, where $V$ is a vector space of dimension $N$. Each component $x_j$ of this vector corresponds to a pixel, that takes the normalized values $x_j \in [0,1]$. Here 0 corresponds to a white pixel and 1 to a black pixel. 

The image vector $x \in V$ is then mapped to a vector $\ket{\Phi(x)}$ in a $2^N$-dimensional vector space $W$, 
\begin{equation}
  W \cong \bigotimes_{j=1}^N W_j,  
\end{equation}
by a transformation $\Phi: V\rightarrow W$ known as the feature map, $\Phi: x \mapsto |\Phi (x) \rangle$. Above, $W_j$ is a 2-dimensional vector space. Following the language of quantum information, we refer to space $W_j$ as a qubit, we call vectors such as $\ket{\Phi(x)}$ ``wavefunctions" or ``states", and we represent them with kets $\ket{~}$. Accordingly, we say that the feature map $\Phi$ maps an image $x$ of $N$ pixels into a state $\ket{\Phi(x)} \in W$ of $N$ qubits. The feature map $\Phi$ is chosen such that the resulting state $\ket{\Phi(x)}$ is normalized to 1 (in $L^2$ norm), i.e. $\langle \Phi(x) | \Phi(x) \rangle = 1$.

The feature map $\Phi$ is also often taken to be comprised of local feature maps $\phi^j$, which are applied to entry $x_j$:
\begin{equation} \label{eq:feature1}
    |\Phi(x)\rangle = \bigotimes_{j=1}^N |\phi^j(x_j)\rangle,~~~~~\ket{\phi^{j}(x_j)} \in W_j.
\end{equation}
That is, each pixel is mapped into a qubit, and the resulting state is called a \textit{product state}, since it can be expressed as a tensor product $\ket{\Phi(x)} = \ket{\phi^1(x_1)} \otimes \ket{\phi^2(x_2)} \otimes \cdots \otimes \ket{\phi^N(x_N)}$.
A typical local feature map is 
\begin{equation} \label{eq:feature2}
    |\phi^j(x_j)\rangle = \cos(\frac{\pi}{2}x_j) |0 \rangle + \sin(\frac{\pi}{2}x_j) |1\rangle,
\end{equation}
where $\{\ket{0}, \ket{1} \}$ is an orthonormal basis, known as the computational basis of the qubit. Notice that this feature map, which acts in the same way across all pixels $j$ of the image, maps white pixels ($x_j=0$) to the $\ket{0}$ state and black pixels ($x_j = 1$) to the $\ket{1}$ state. 

For ease of notation, in the rest of this paper we write $\ket{x}$ to mean the state $\ket{\Phi(x)}$. After the feature map has been applied, images are classified as follows. Let $\{ \ket{T_\ell} \}$ denote a set of N-qubit variational states encoded in a tensor network model, where state $\ket{T_\ell} \in W$, and the index $\ell$ is a label for the classes under consideration. For instance, $\ell \in \{0,1,...,9\}$ for the MNIST data set of handwritten digits. Given an image $x$ encoded in the state $|x\rangle$, $x$ is classified as the label $k$ for which the overlap $|\langle T_k | x \rangle |$ is largest:
\begin{equation} \label{eq:classification}
    k = \mathrm{argmax}_\ell |\langle T_\ell | x \rangle |.
\end{equation}

This model is then trained by choosing the variational parameters in the tensor network such that some loss function is minimized on the training set $\mathcal{T} = \{(x^{(i)}, y^{(i)})\}_{i=1}^{N_T}$. Here, $x^{(i)}$ are the images in the training set, and $y^{(i)}$ are the corresponding correct labels for these images (i.e. the train labels), whereas $N_T$ denotes the number of images in the training set.  Previous studies employed the quadratic loss function
\begin{equation} \label{eq:loss}
    J \big( \{\ket{T_\ell} \} \big) = \frac{1}{2} \sum_{i=1}^{N_T} \sum_{\ell} \Big( \big| \big\langle T_\ell \big| x^{(i)} \big\rangle \big| - \delta_{\ell,y^{(i)}} \Big)^2,
\end{equation}
where $\delta_{\ell,y^{(i)}}$ is the Kronecker delta. This loss function penalizes the difference between $\left|\braket{ T_\ell}{ x^{(i)}} \rangle\right|$ and its ideal output, $\delta_{\ell,y^{(i)}}$ (1 for the correct label, and 0 otherwise).

Finally, once the tensor network model has been trained using the training set, it is tested by applying the feature map $\Phi$ to the images in the test set and by then classifying them using Eq. \eqref{eq:classification}.

Given the feature map in Eqs.~\eqref{eq:feature1}-\eqref{eq:feature2} and the loss function in Eq.~\eqref{eq:loss}, the performance of the tensor network model still depends critically on which specific tensor network we use in order to encode the variational states $\ket{T_{\ell}}$. Let us consider three examples:

\begin{itemize} 
\item Product state: the simplest possible tensor network model corresponds to a product state, $\ket{T_{\ell}} = \ket{\Psi^{\PS}_{\ell}}$, where  $\ket{\Psi^{\PS}_{\ell}} = \bigotimes_{j=1}^N \ket{\psi^{j}_{\ell}}$ specifies a different state $\ket{\psi^{j}_{\ell}} \in W_{j}$ for each of the $N$ qubits. Since the state of each qubit can be specified with 2 parameters, we can specify the product state $\ket{\Psi^{\PS}_{\ell}} \in W$ using $2N$ parameters.
\item Generic state: In the opposite extreme, the most complicated tensor network model would be to not restrict the $N$-qubit state at all, but consider instead a generic state, $\ket{T_{\ell}} = \ket{\Psi^{\generic}_{\ell}}$, where $\ket{\Psi^{\generic}_{\ell}} \in W$ is specified by $2^N$ parameters.
\item Matrix product state (MPS): In between, one finds the MPS, $\ket{T_{\ell}} = \ket{\Psi^{\MPS}_{\ell}}$, as used in Ref. \cite{2016arXiv160505775M}. The MPS $\ket{\Psi^{\MPS}_{\ell}} \in W$ is specified by $O(\chi^2 N)$ parameters, if it is constructed from $\chi \times \chi$ matrices.
\end{itemize}

In terms of expressive power and computational costs, the product state is the least expressive and least expensive, with computational memory and time (per sample) scaling as $O(N)$. In contrast, a generic state $\ket{\Psi^{\generic}_{\ell}}$ is the most expressive (it can express any linear map in $W$!). However, storing and using a generic state $\ket{\Psi^{\generic}_{\ell}}$ incurs computational memory and time that grows exponentially in $N$, and it is thus not an affordable option for large $N$. Finally, the MPS $\ket{\Psi^{\MPS}_{\ell}}$ sits between the previous two options. It is more expressive than a product state but less so than a generic state, and it has computational cost $O(\chi^2 N)$ (per sample). 

One thus finds a trade-off between expressive power and computational efficiency depending on the complexity of the tensor network. While a tensor network model needs to be both sufficiently expressive and computationally efficient for a given task, generalization is yet also another very important property to take into consideration. Avoiding over-fitting in order to achieve sufficient generalization relates not only to the model, but also to how it is optimized, making a systematic analysis much more difficult. 

\subsection{Modified Approach} \label{sec:OurApp}

In this work we employ a variation of the protocol outlined above. We use the same local feature map $\Phi$ in Eq. \eqref{eq:feature2} to encode the image data $x$ into a product state $\ket{x}$. Given a tensor network model that produces a state $\ket{T_{\ell}}$ for each class $\ell$, we also use the same classification criterion in Eq. \eqref{eq:classification}. However, we use a different loss function. 

Let $\ket{x^{(i)}}$ be the state corresponding to embedding the training image $x^{(i)}$, and let us first define a probability distribution (inspired by the so-called Born rule of quantum mechanics) given by 
\begin{equation}
    p\left( y^{(i)}=\ell \right) \equiv  \frac{|\langle x^{(i)}| T_{\ell} \rangle |^2}{\sum_k |\langle x^{(i)}| T_k \rangle |^2}.
\end{equation}
Notice that, indeed, this is a probability distribution since by construction we have
\begin{equation}
p\left(y^{(i)}=\ell\right) \geq 0,~~~ \sum_{\ell} p\left(y^{(i)} = \ell \right) = 1.
\end{equation}
Notice also that we can replace the classification criterion \eqref{eq:classification} with the equivalent classification criterion
\begin{equation} \label{eq:classification2}
    k = \mathrm{argmax}_\ell ~p\left( y^{(i)}=\ell \right).
\end{equation}
Then, instead of optimizing a quadratic loss, we optimize the negative log-likelihood:
\begin{equation}
    J\big( \{ |T_\ell \rangle \} \big) = -\sum_{i=1}^{N_T} \sum_\ell \delta_{\ell, y^{(i)}} \log\big(p(y^{(i)} = \ell )\big).
\end{equation} 
This loss function is minimized when it perfectly classifies the training set, namely when we have $p(y^{(i)}=\ell) = \delta_{\ell, y^{(i)}}$. Notice that our loss is similar to the loss functions used in Refs. \cite{2019arXiv190606329E, 2020arXiv200410076S}. However, here we work with the logarithm of the overlap, instead of the (logarithm of the exponential of the) overlap. Our formulation is better prepared to deal with overlaps in an $N$-qubit Hilbert space, which are exponentially large (or small) in
$N$. 

Finally, another important difference is that instead of using a MPS $\ket{\Psi^{\MPS}_{\ell}}$ as in Ref. \cite{2016arXiv160505775M}, here we will explore the use of a simpler tensor network, representing a block product state $\ket{\Psi^{\BPS}_{\ell}}$, as described in Sect. \ref{sec:IV}. 
 
\section{Entanglement and image classification} \label{sec:III}

In this section we investigate the entanglement structure of a state $\ket{\Sigma_{\ell}} \in W$ that we initially thought might be closely related to what a tensor network model might be trying to learn. We will conclude, however, that $\ket{\Sigma_{\ell}}$ is too entangled for the tensor network model in Ref. \cite{2016arXiv160505775M} to learn it, even approximately.

Specifically, for each label $\ell$ we consider the state 
\begin{equation}
        \ket{\Sigma_\ell} \equiv \sum_{i;\ y^{(i)}=\ell } \ket{x^{(i)}},
\label{eq:Sigma_l}    
\end{equation}
that is, a linear combination of all the states $\ket{x^{(i)}}$ corresponding to images $x^{(i)}$ in the training set that are classified in class $\ell$. Note that this state is not normalized.

By construction, this state has significant overlap with any image in the training set that is labelled $\ell$. Indeed, for such images $\langle x^{(i)} \ket{\Sigma_\ell} \geq 1$. Hence, using $\ket{\Sigma_\ell}$ for classification yields reasonable accuracies on the training set; it was also observed to produce reasonable accuracies on the test set. May it then be the case that the MPS model $\ket{\Psi^{\MPS}_\ell}$ in Ref. \cite{2016arXiv160505775M} somehow approaches $\ket{\Sigma_\ell}$ during training? To address this question, next we study the entanglement structure of $\ket{\Sigma_\ell}$, and we compare it to the entanglement structure allowed in an MPS.
 
\subsection{Schmidt rank and entanglement entropy}

Let us partition the $N=28 \times 28$ qubits into two sets $A$ and $B$, where $A$ will be some subset of adjacent qubits to be described below. Let us define a normalized version $|\hat{\Sigma}_{\ell}\rangle$ of state $\ket{\Sigma_{\ell}}$, that is
\begin{equation}
    |\hat{\Sigma}_{\ell}\rangle \equiv \frac{\ket{\Sigma_{\ell}}}{\sqrt{\braket{\Sigma_{\ell}}{\Sigma_{\ell}}}}, 
\end{equation}
and then expand it in its Schmidt decomposition
\begin{equation}
    |\hat{\Sigma}_{\ell}\rangle = \sum_{\alpha=1}^{\chi} \lambda_{\alpha}\ket{\varphi^A_{\alpha}}\ket{\varphi^B_{\alpha}}.
\end{equation}
Here $\{\lambda_{\alpha}\}_{\alpha=1}^{\chi}$ are the (non-vanishing) Schmidt coefficients, which are sorted in decreasing order, namely $\lambda_{\alpha} \geq \lambda_{\alpha+1} \geq 0$, and fulfill $\langle \hat{\Sigma}_{\ell}|\hat{\Sigma}_{\ell} \rangle = \sum_{\alpha} (\lambda_{\alpha})^2 =1$. In turn, the states $\{\ket{\varphi^{A}_{\alpha}}\}$ form an orthonormal basis, $\braket{\varphi^A_{\alpha}}{\varphi^{A}_{\alpha'}} = \delta_{\alpha\alpha'}$, and the same applies to $\{\ket{\varphi^{B}_{\alpha}}\}$, with $\braket{\varphi^B_{\alpha}}{\varphi^{B}_{\alpha'}} = \delta_{\alpha\alpha'}$.

\begin{figure}[H]
\centering
\includegraphics[width=.4\textwidth]{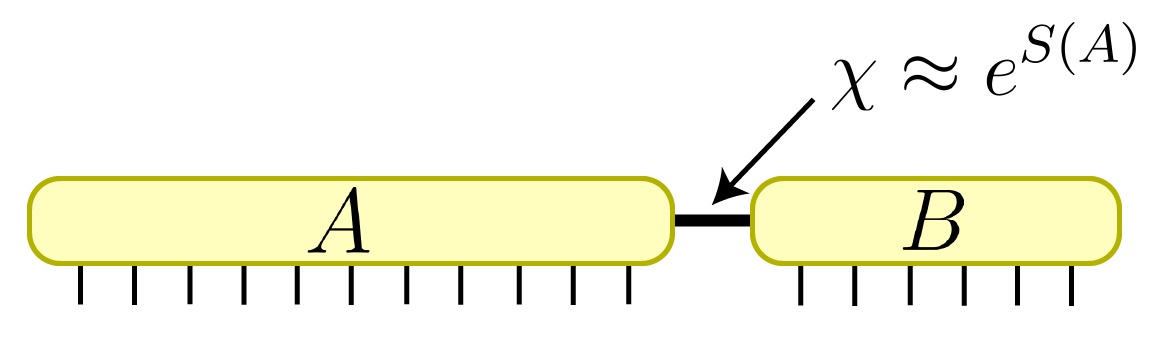}
\caption{A partition of a system into regions $A$ and $B$. In order to represent this system by a tensor network, the bond dimension in between regions $A$ and $B$ must be $\chi \approx e^{S(A)}$.}
\label{fig:AB_partition}
\end{figure}

In order to characterize the entanglement in the above state, we consider two quantities. The first one is the entanglement entropy $S(A)$ (equivalently, $S(B)$) of the state $|\hat{\Sigma}_{\ell}\rangle$ with respect to the partition $A$:$B$, which is defined as
\begin{equation} \label{eq:Schmidt}
    S(A) \equiv - \sum_{\alpha=1}^{\chi} (\lambda_{\alpha})^2 \log\left( (\lambda_{\alpha})^2 \right),
\end{equation}
and it is a measure of how much correlation there is between parts $A$ and $B$. For our purposes, the entanglement entropy provides a useful lower bound, namely $e^{S(A)}$, on the minimal bond dimension that needs to be connecting parts $A$ and $B$ in a tensor network representation of $|\hat{\Sigma}_{\ell}\rangle$, see Fig. \ref{fig:AB_partition}. 

A more direct measure of the required bond dimension is given by a second quantity, the Schmidt rank $\chi$, that is, the number of non-vanishing Schmidt terms in the decomposition \eqref{eq:Schmidt}. When all the Schmidt coefficients are of similar size, then the Schmidt rank $\chi$ is a robust measure of the bond dimension needed in a tensor network that accurately approximates the state $|\hat{\Sigma}_{\ell}\rangle$, and we have $\chi \approx e^{S(A)}$. However, if the Schmidt coefficients have very different sizes, then it might be possible to truncate (ignore) some of the terms in the Schmidt decomposition corresponding to the smallest Schmidt coefficients while still obtaining an accurate approximation of the state $|\hat{\Sigma}_{\ell}\rangle$, in which case a total bond dimension smaller than $\chi$ may already be sufficient in an approximate tensor network representation of $|\hat{\Sigma}_{\ell}\rangle$. Below we report results for $|\hat{\Sigma}_{\ell = 3}\rangle$, that is, for MNIST images of the digit `3', although the same construction for other values of the class label $\ell \in \{0,1,...,9\}$ produces very similar results.

\subsection{Partition into top and bottom halves}

\begin{figure}[H]
\centering
\includegraphics[width=.44\textwidth]{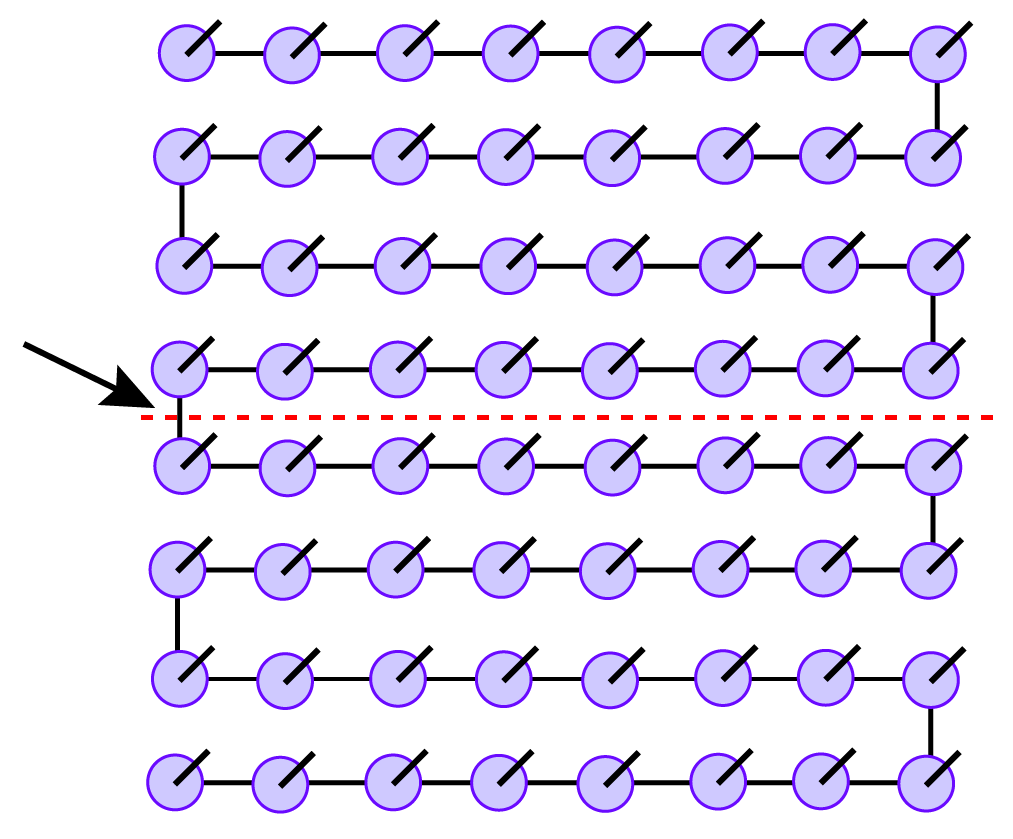}
\caption{Example of an MPS that snakes around a two-dimensional square lattice of qubits, used to encode images of $8 \times 8$ pixels. The discontinuous line partitions the top and bottom halves of the image. The MPS only has one bond index, emphasized with an arrow, connecting the top and bottom halves.}
\label{fig:MPS_Half}
\end{figure}

In Ref. \cite{2016arXiv160505775M}, the MPS \textit{snakes} around the $28\times 28$ square lattice of qubits (which had been reduced to a $14 \times 14$ square lattice of qubits for simplicity) by moving from left to right, then right to left, and so on, while descending through the grid, see Fig. \ref{fig:MPS_Half}. That means that the top half $A$ and bottom half $B$ of the lattice, each made of $14\times 28 = 392$ qubits, are only connected by one single bond index. In Ref. \cite{2016arXiv160505775M}, this bond index was chosen to take up to 120 values, in which case the classification task had test accuracy of $99.03\%$.

\begin{figure}[H]
\centering
\includegraphics[width=.47\textwidth]{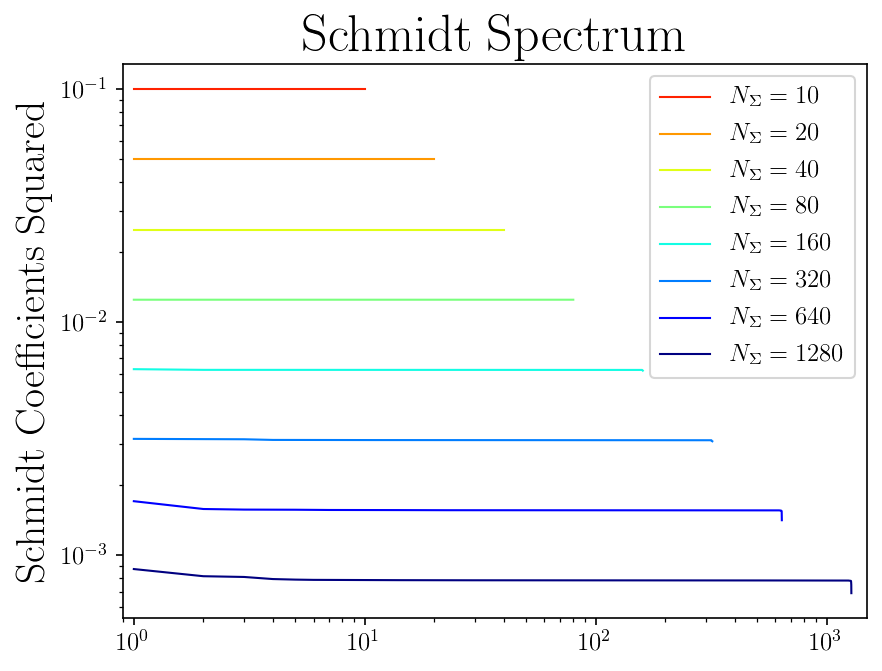}
\caption{Schmidt spectrum for different $N_\Sigma$ in the range $10-1280$ of the state $\ket{\Sigma_3}$, constructed from encoded MNIST images of the digit `3'. Part $A$ is the top half of the square lattice of qubits. (Notice that we plot $(\lambda_{\alpha})^2$ instead of $\lambda_{\alpha}$). For small $N_\Sigma$ the lines are horizontal, that is, all the Schmidt values have essentially the same magnitude $\lambda_{\alpha} \approx 1/\sqrt{N_{\Sigma}}$. 
}
\label{fig:Spectrum_Half}
\end{figure}

Fig. \ref{fig:Spectrum_Half} shows the Schmidt spectrum of this partition, as a function of the total number $N_\Sigma$ of images used in the training set, for $|\hat{\Sigma}_3 \rangle$ -- that is, for images corresponding to the digit $3$. We find that the Schmidt spectrum is essentially flat, indicating that the required bond dimension for an accurate MPS description of $|\hat{\Sigma}_3 \rangle$ is essentially equal to $N_\Sigma$. For instance, for $N_\Sigma = 1280$ images, the maximal bond dimension $120$ used in Ref. \cite{2016arXiv160505775M} results in an MPS that cannot be, even by far, an accurate approximation to $|\hat{\Sigma}_3 \rangle$, because $120 \ll 1280$. We conclude that the MPS in Ref. \cite{2016arXiv160505775M}, which successfully classifies the images, is not representing a state anywhere close to $|\hat{\Sigma}_{3}\rangle$.

A flat spectrum of Schmidt values in $|\hat{\Sigma}_3 \rangle$ indicates that the bottom (and top) of the $N_\Sigma$ images in the training set are encoded in essentially orthonormal states. That follows simply from the fact that any two images typically differ in a few number of pixels both on the top half and on the bottom half. For larger values of $N_\Sigma$ we see that the Schmidt values are no longer the same, although they are still very similar. This indicates that some of the images in the training set are now a bit similar, in that their overlaps in the top or bottom halves are no longer negligible. However, an accurate approximation to $|\hat{\Sigma}_3 \rangle$ still requires keeping about $N_\Sigma$ of the Schmidt values, so that an MPS representing  $|\hat{\Sigma}_3 \rangle$ (even approximately) would need to have bond dimension $\approx N_\Sigma$.  

In the case of a flat spectrum $\lambda_{\alpha} \approx 1/\sqrt{N_\Sigma}$, the entanglement entropy is given by $S(A) \approx \log N_\Sigma$. Since the spectrum in Fig. \eqref{fig:Spectrum_Half} is very flat, here we do not learn anything new by studying at the entanglement entropy (not plotted), but since this is the most popular measure of entanglement, we include reference to it to facilitate comparison with other research. 

Finally, we point out that computing the Schmidt decomposition of $|\hat{\Sigma}_{\ell}\rangle$ in vector spaces of very large dimension (notice that $28\times 28 = 784$ qubits are described by a vector space $W$ of dimension $2^{784} \approx 10^{236}$) can be accomplished with computational cost $O((N_{\Sigma})^{3})$ using the strategy described in the Appendix.

\subsection{Central block of size $L\times L$}

For completeness, we have also explored the amount of entanglement entropy of a square region A of size $L\times L$. Specifically, for $|\hat{\Sigma}_3 \rangle$ we computed the average entropy of a square of $L\times L$ qubits in a central window of size $10 \times 10$. For instance, when $L=1$, we looked at the average entropy of all 100 qubits in this central window; when $L=2$, we looked at the average entropy of all 81 $2 \times 2$ squares of qubits in this window; and so on. 

\begin{figure}[h!]
\center{\includegraphics[width=.47\textwidth]{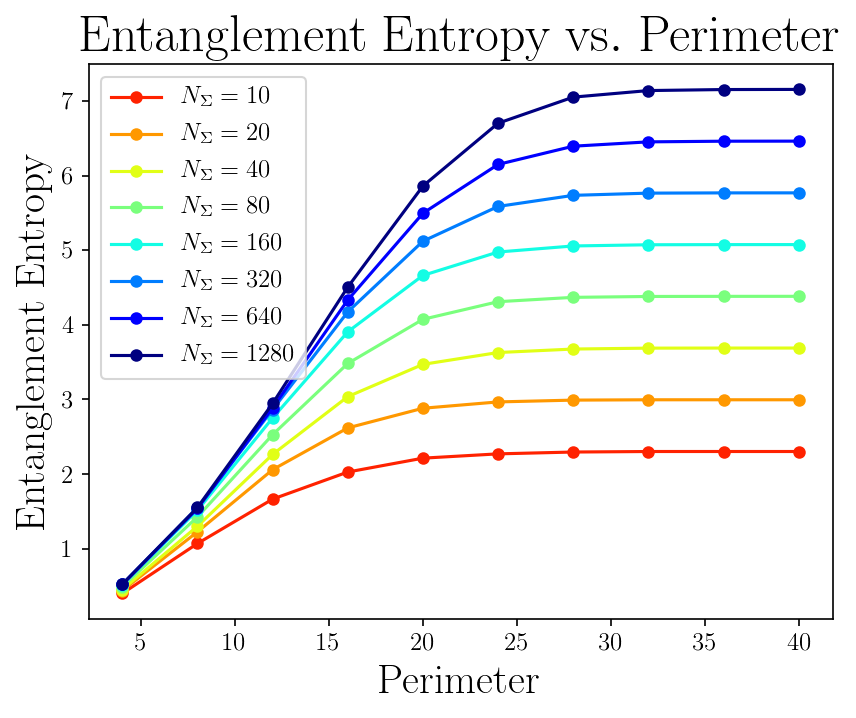}}
\caption{Average entanglement entropy $S(A)$ vs. perimeter $4L$ for regions $A$ consisting of $L\times L$ squares in a $10\times 10$ central window, for the state $|\hat{\Sigma}_3 \rangle$ constructed from encoded MNIST images of the digit `3'.}
\label{fig:EntropyPlot}
\end{figure}

We display our results in Figure \ref{fig:EntropyPlot} below for states $|\hat{\Sigma}_3 \rangle$ built as a superposition of $N_\Sigma$ images, for a range of values of $N_\Sigma$. We see that for a block of size $L\times L$, the entropy appears to grow (slighly faster than) linearly in the perimeter size $4L$, before saturating very close to its maximal possible value for $N_\Sigma$ images, namely $\log(N_\Sigma)$. 
  
\begin{figure}[h!]
\center{\includegraphics[width=.47\textwidth]{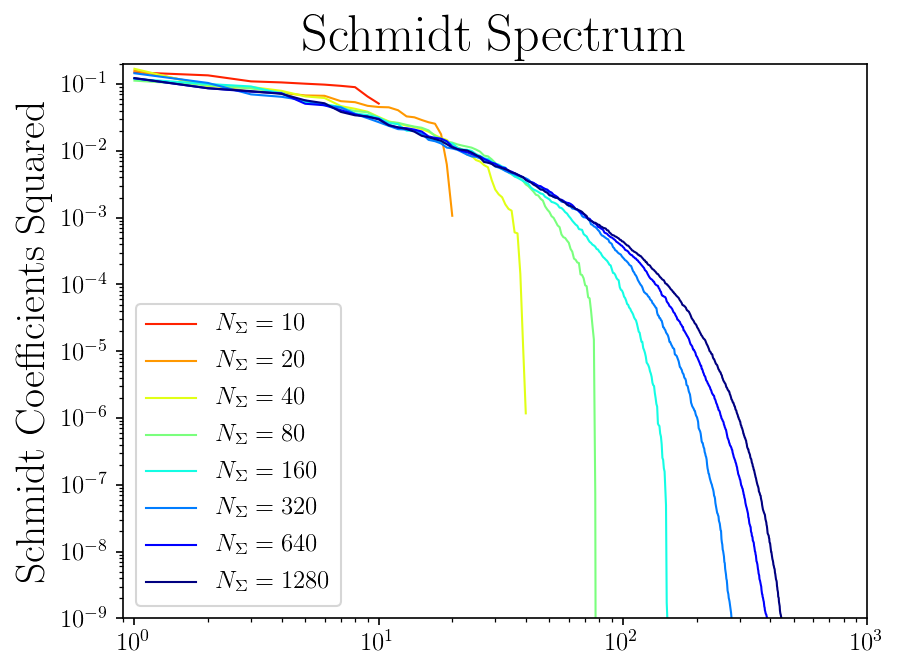}}
\caption{Schmidt spectrum $\{\lambda_{\alpha}\}$ for different values of $N_\Sigma$, when region $A$ is a $3\times3$ square in the central window of $|\hat{\Sigma}_3 \rangle$, constructed from encoded MNIST images of the digit `3'. (Notice that we plot $(\lambda_{\alpha})^2$ instead of $\lambda_{\alpha}$.)}
\label{fig:Spectrum_3x3}
\end{figure}

To gain further insight, Figure \ref{fig:Spectrum_3x3} shows the Schmidt spectrum in the case where part $A$ is a square block of $3\times 3$ qubits, again as a function of $N_\Sigma$. Notice that the vector space of $3\times 3 = 9$ qubits has dimension $2^9=512$, which provides an upper bound for the Schmidt rank of $|\hat{\Sigma}_3 \rangle$ with respect to this partition. When $N_\Sigma=10$, we observe a rather flat Schmidt spectrum, indicating that the $N_\Sigma$ images are embedded in fairly orthogonal states both in $A$ and its complement $B$. However, as the number $N_\Sigma$ of images grows, the corresponding states in region $A$ start to overlap non-trivially, and this results in a sharply decaying spectrum of Schmidt values, whose magnitude is seen to range e.g. from $10^{-1}$ to $10^{-9}$. This indicates that one could in principle truncate away the terms in the Schmidt decomposition corresponding to the smallest Schmidt values while retaining an accurate approximation to $|\hat{\Sigma}_{3}\rangle$. However, the number of Schmidt values one needs to keep is seen to grow sharply with $L$, as indicated by the entanglement entropy in Figure \ref{fig:EntropyPlot}. This implies that a tensor network such as MPS or tree tensor network would require a very large bond dimension to represent $|\hat{\Sigma}_3 \rangle$, making such representation inefficient.

\section{Expressive power of block product states} \label{sec:IV}

In the previous section we have seen that the state $\ket{\Sigma_{\ell}}$ in Eq. \eqref{eq:Sigma_l}, built by simply superposing the encoded images of class $\ell$ in the training set, was very robustly entangled, so much so that it precluded an efficient representation in terms of the MPS used in Ref. \cite{2016arXiv160505775M} to successfully classify this data set. We concluded that a tensor network such as an MPS does not need to be able to represent the state $\ket{\Sigma_{\ell}}$ in order to be a successful model for image classification. 

With this insight, we next explore the use of other simple tensor network models for the same task. Specifically, we will consider tensor networks that represent states with entanglement restricted within small blocks of qubits. We will learn that these simple tensor networks are already very expressive. However, we will also see that, at least with our current optimization algorithm, these models suffer from over-fitting and therefore generalize poorly from the training data set to the test data set. We will then investigate ways to alleviate this problem, with partial success, and will conclude that further research is still needed to prevent over-fitting in these otherwise quite promising, surprisingly simple tensor network models.

 \subsection{Block Product States} \label{sec:BPS}

\begin{figure}[h!]
\center{\includegraphics[width=.47\textwidth]{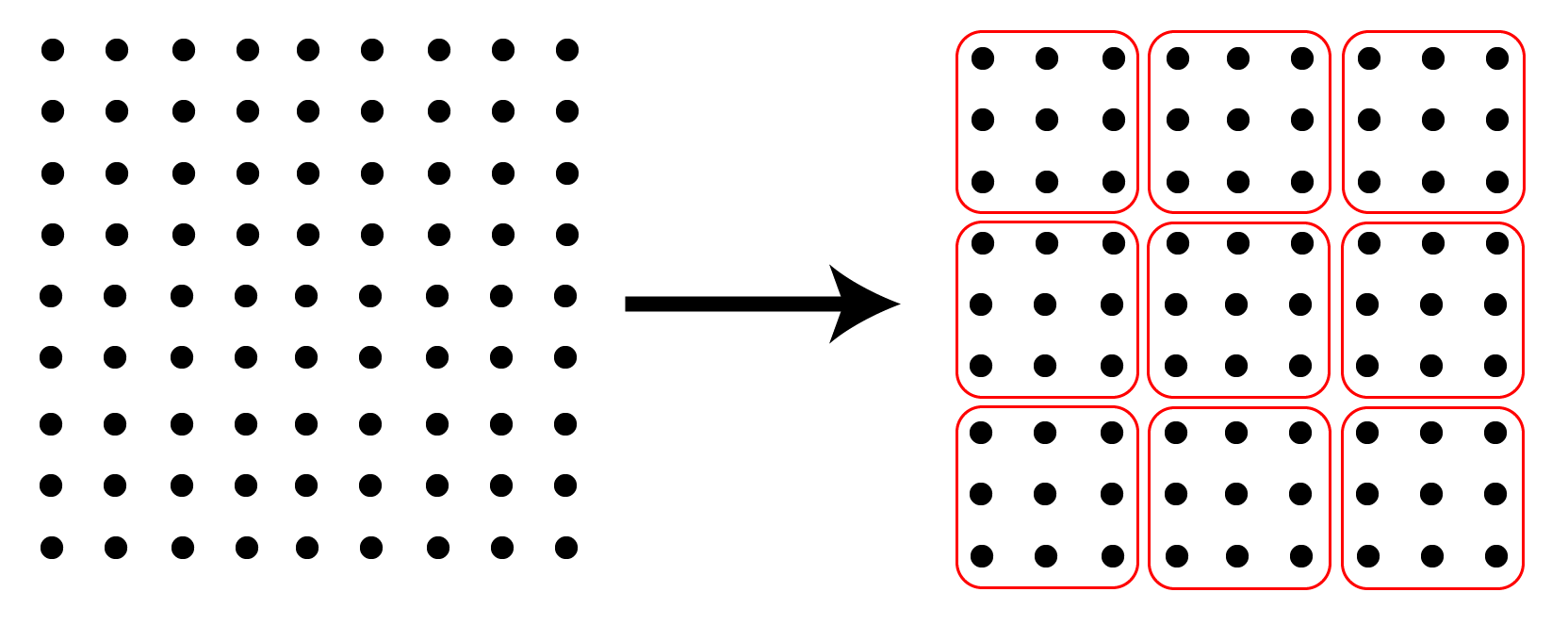}}
\caption{Tiling a grid into blocks of size $n\times n$, where $n=3$.}
\label{fig:Blocks}
\end{figure}

\begin{figure}[h!]
\center{\includegraphics[width=.47\textwidth]{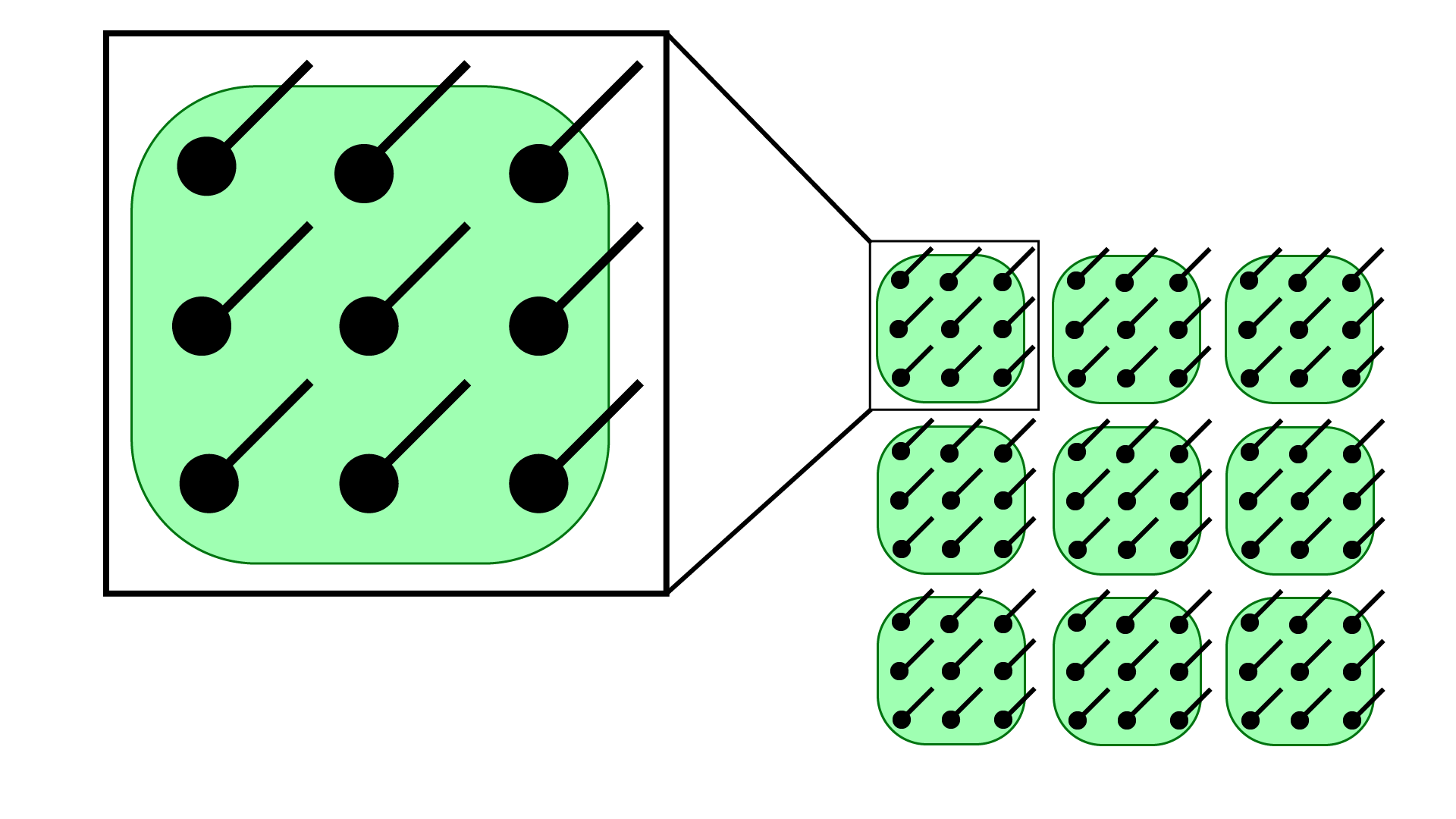}}
\caption{Construction of a block product state from the set of blocks. Here, $n=3$.}
\label{fig:BPS}
\end{figure}

We first define the general structure of the states used in the following models. Given the square lattice of $28 \times 28$ qubits in which the MNIST images have been encoded, we consider subdivisions into square blocks of $n\times n$ adjacent qubits for $n=1,2,3,4$, see Fig. \ref{fig:Blocks} for an illustration with $n=3$. For $n=1,2,3$ and $4$, we respectively obtain $28^2$, $14^2$, $9^2$ and $7^2$ such blocks; for $n=3$, we ignored the last row and column of pixels (nearly all of which are black anyway) so that the images were encoded in a square lattice made of $27\times 27$ qubits. We then take the tensor network state $\ket{T_\ell}$ to be a ``block product state" $\ket{\Psi^{\BPS}_{\ell}}$, namely a state that can be written as the tensor product of states $\ket{\psi_{\ell}^{b}}$ for each square block $b$ of $n\times n$ qubits, that is
\begin{equation}
    \ket{\Psi^{\BPS}_{\ell}} \equiv \bigotimes_{b\in B_n} |\psi_{\ell}^{b}\rangle,
\end{equation}
A block product state is represented diagrammatically in Fig. \ref{fig:BPS}.
Notice that $\ket{\psi_b}$ is itself a state of $n^2$ qubits. Its number $d=2^{n^2}$ of components grows very fast with $n$. Indeed, for $n=1,2,3$, and $4$ it is $d=2, 16, 512$ and $65,\!536$, respectively. We will then further specialize the block product state structure, by replacing each generic state $\ket{\psi_b}$ made of $d=2^{n^2}$ components with a more efficient tensor network representation. Below we consider two options: the nearest neighbor block product state, which consists of a projected-entangled pair state PEPS \cite{2004cond.mat..7066V} within each $n\times n$ block, and the snake block product state, which is an MPS within each $n\times n$ block, as described below.

\subsection{Nearest Neighbor Block Product State}

\begin{figure}[h!]
\center{\includegraphics[width=.47\textwidth]{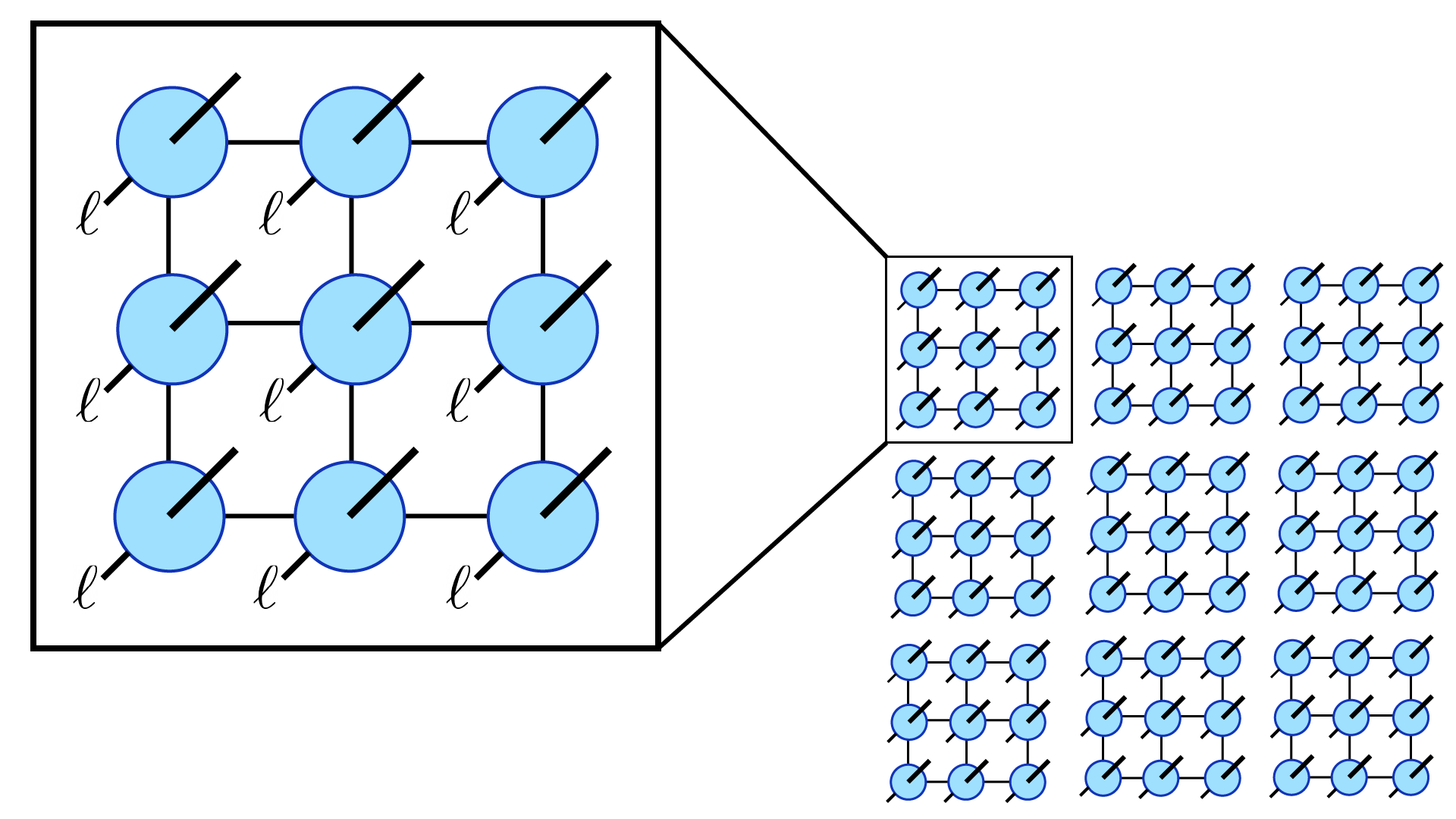}}
\caption{A nearest neighbor block product state (NNBPS).}
\label{fig:BPS_PEPS}
\end{figure}

Fig. \ref{fig:BPS_PEPS} depicts a \textit{nearest neighbor block product state} (NNBPS), in which the state $\ket{\psi_{\ell}^{b}}$ for block $b \in B_n$ is represented by a PEPS, where each PEPS tensor has bond indices connecting it to its nearest neigbor tensors within the $n\times n$ block. We choose the bond dimension $\chi=2$, so that a PEPS tensor with 4 bond indices and one pixed index consists of $2^5=32$ parameters. Notice that we also endow each tensor with a class label $\ell$.  

To train the model, we minimize the loss function outlined in Sec. \ref{sec:OurApp} with the Adam optimization algorithm. In addition, we include in the loss function a regularization term to keep the normalization of $\{|T_\ell \rangle \}$ finite: $+\ \alpha \sum_\ell |\log(Z_\ell)|$, where $Z_\ell = \langle T_\ell | T_\ell \rangle$. In our analyses, we let $\alpha \sim O(1)$. We display results below in Tables \ref{tab:MNIST_PEPS} and \ref{tab:Fashion_PEPS}.

\begin{table}[h!]
    \begin{tabular}{ |c||c|c| } 
    \hline 
    Block Size & Training Accuracy & Test Accuracy \\
    \hline
    \hline
    $1 \times 1$ & 93.070\% & 91.100\% \\
    \hline
    $2 \times 2$ & 99.967\% & 94.690\% \\
    \hline
    $3 \times 3$ & 99.925\% & 95.470\% \\ 
    \hline
    $4 \times 4$ & 99.977\% & 95.420\% \\ 
    \hline
    \end{tabular}
    \caption{Nearest neighbor block product state applied to MNIST dataset of handwritten digits}
    \label{tab:MNIST_PEPS}
\end{table}

\begin{table}[h!]
    \begin{tabular}{ |c||c|c| } 
    \hline 
    Block Size & Training Accuracy & Test Accuracy \\
    \hline
    \hline
    $1 \times 1$ & 88.132\% & 84.230\% \\
    \hline
    $2 \times 2$ & 92.788\% & 86.540\% \\
    \hline
    $3 \times 3$ & 94.275\% & 86.890\% \\ 
    \hline
    $4 \times 4$ & 94.940\% & 87.320\% \\ 
    \hline
    \end{tabular}
    \caption{Nearest neighbor block product state applied to Fashion-MNIST dataset}
    \label{tab:Fashion_PEPS}
\end{table}

On the MNIST dataset of handwritten digits, we see that even small $2 \times 2$ blocks can achieve nearly 100\% training accuracy. We find that rather remarkable. It means that such a simple tensor network model already has the \textit{potential} of being able to classify also the MNIST images in the test set with the same accuracy (after all, this is what would happen if we included the  test set in the training set). As it is well-known, however, having enough expressive power to classify all the images is only useful if we also know how to train the model, using only the training set, in a way that it suitably generalizes to the test set. And this is where our approach still fails. For a $2\times 2$ block, our current optimization scheme results in poor test accuracies, under $95\%$. Blocks of size $3\times 3$ and $4\times 4$ are seen to again lead to nearly $100\%$ train accuracies but much lower test accuracies under $96\%$.

We have also explored performance on the Fashion-MNIST dataset. We found that train and test accuracies monotonically increase with block size, but again the test accuracy lags behind the training accuracy significantly. In addition, as this data set is more complex than MNIST digits, we do not achieve 100\% training accuracy, while the test accuracy saturates around $\sim 87\%$. We note nevertheless that this accuracy is comparable to that of Ref. \cite{2019arXiv190606329E}, where $88\%$ test accuracy was achieved on the Fashion-MNIST data set using an MPS model.

We conclude that this first block product state model is, surprisingly, expressive enough to fit the training set very well, but clearly over-fits the data.

\subsection{Snake Block Product State}

In an attempt to reduce over-fitting, we have explored the use of alternative tensor networks to represent the state $\ket{\psi_{\ell}^b}$ within each block. Here we report on one of them, which for a block of size $4\times 4$ resulted in lower training accuracy but higher test accuracy than the NNBPS described above.

\begin{figure}[h!]
\center{\includegraphics[width=.47\textwidth]{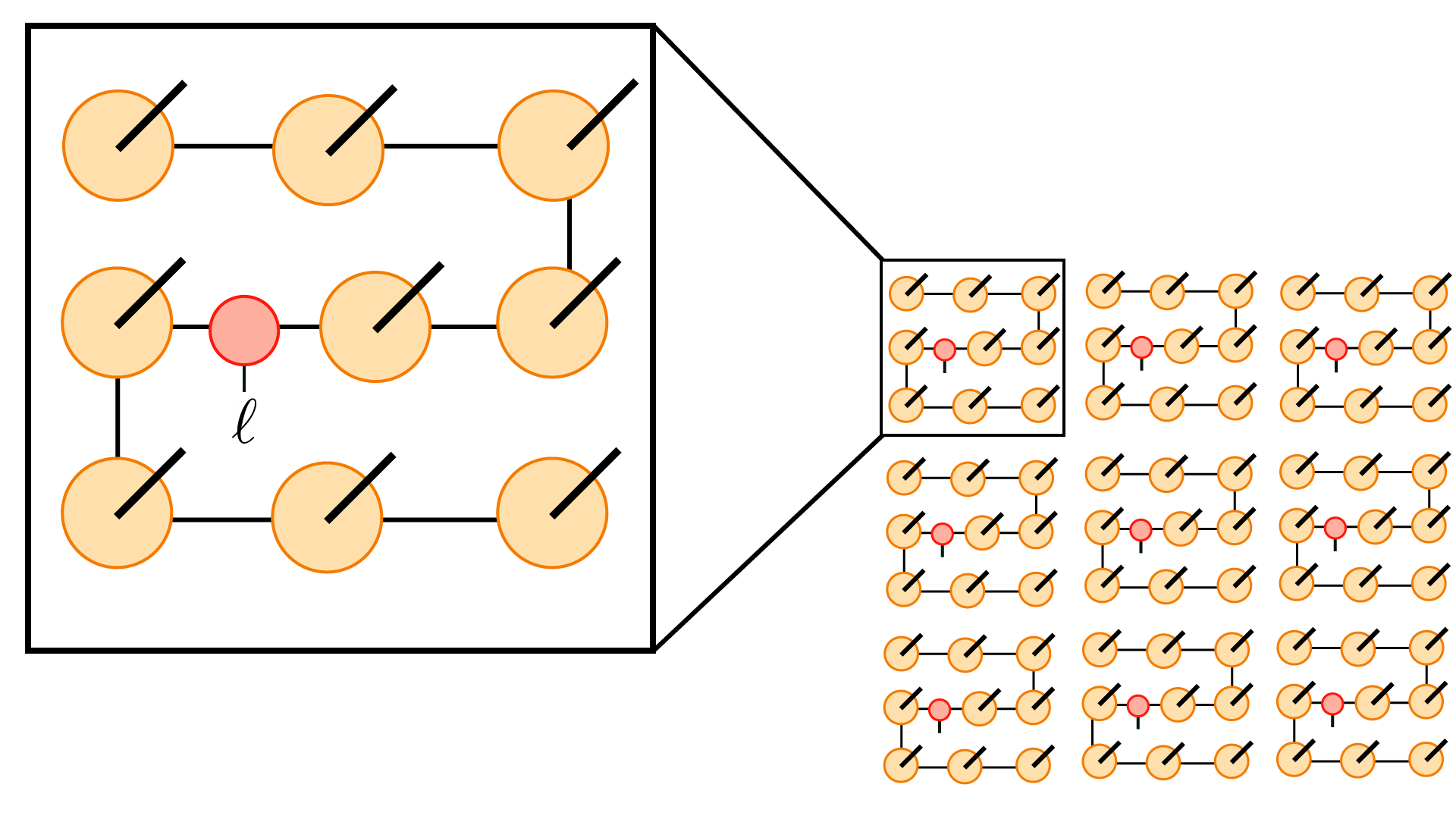}}
\caption{A snake block product state (SBPS).}
\label{fig:BPS_MPS}
\end{figure}

Fig. \ref{fig:BPS_MPS} depicts a \textit{snake block product state} (SBPS), in which the state $\ket{\psi_{\ell}^{b}}$ for block $b \in B_n$ is represented by an MPS with its bond index scanning the $n\times n$ block by moving from left to right in the top row, then right to left in the next row, etc, imitating a snake. We consider blocks of size $n\times n$ for $n=2,3,4$ (notice that the case $n=1$ would be identical to the previous analysis). In addition, in order to reduce variational parameters and/or frustrate their optimization, we only have one $\ell$ label for each MPS, which hangs from an additional tensor connected to the MPS tensors through two bond indices, see Fig. \eqref{fig:BPS_MPS}. Using a single class label $\ell$ for the whole MPS (as opposed to having a class label $\ell$ on each tensor of the MPS) seems to help lower the training accuracy while lifting the test accuracy. This may be due to the fact that the parameters in the rest of the MPS tensors are shared among the different classes. (We also implemented the same `single class label' on each PEPS of the NNBPS described above, but in that case we did not obtain better results.)

We train the SBPS model by again minimizing the loss function outlined in Sec. \ref{sec:OurApp} with Adam optimization and add the same regularization term before to prevent normalization problems. We choose $\alpha \sim O(1)$ and vary the bond dimension of the network between $\chi=2$ and $\chi=18$. We display results below in Table \ref{tab:MNIST_MPS}.

\begin{table}[H]

    \begin{tabular}{ |c|c||c|c| } 
    \hline 
    Block Size & Bond Dim. & Training Accuracy & Test Accuracy \\
    \hline
    \hline
    $2 \times 2$ & $\chi=2$ & 96.000\% & 94.700\% \\
    \hline
    $2 \times 2$ & $\chi=3$ & 97.048\% & 95.330\% \\
    \hline
    $2 \times 2$ & $\chi=4$ & 97.983\% & 95.710\% \\
    \hline
    \end{tabular}

    \vspace*{.5 cm}
    
    \begin{tabular}{ |c|c||c|c| } 
    \hline 
    Block Size & Bond Dim. & Training Accuracy & Test Accuracy \\
    \hline
    \hline
    $3 \times 3$ & $\chi=2$ & 94.598\% & 94.000\% \\
    \hline
    $3 \times 3$ & $\chi=3$ & 96.757\% & 95.130\% \\
    \hline
    $3 \times 3$ & $\chi=4$ & 97.657\% & 95.890\% \\
    \hline
    $3 \times 3$ & $\chi=6$ & 97.808\% & 95.640\% \\
    \hline
    $3 \times 3$ & $\chi=12$ & 98.367\% & 95.430\% \\
    \hline
    $3 \times 3$ & $\chi=18$ & 98.085\% & 95.470\% \\
    \hline
    \end{tabular}
    
    \vspace*{.5 cm}
    
    \begin{tabular}{ |c|c||c|c| } 
    \hline 
    Block Size & Bond Dim. & Training Accuracy & Test Accuracy \\
    \hline
    \hline
    $4 \times 4$ & $\chi=2$ & 94.003\% & 93.870\% \\
    \hline
    $4 \times 4$ & $\chi=3$ & 96.342\% & 95.300\% \\
    \hline
    $4 \times 4$ & $\chi=4$ & 96.820\% & 95.390\% \\
    \hline
    $4 \times 4$ & $\chi=6$ & 97.878\% & 96.200\% \\
    \hline
    $4 \times 4$ & $\chi=12$ & 97.050\% & 95.340\% \\
    \hline
    $4 \times 4$ & $\chi=18$ & 97.332\% & 94.550\% \\
    \hline
    \end{tabular}
    \caption{Block product state constructed from MPS applied to the MNIST dataset of handwritten digits}
    \label{tab:MNIST_MPS}
\end{table}

\begin{figure}[h!]
\center{\includegraphics[width=.47\textwidth]{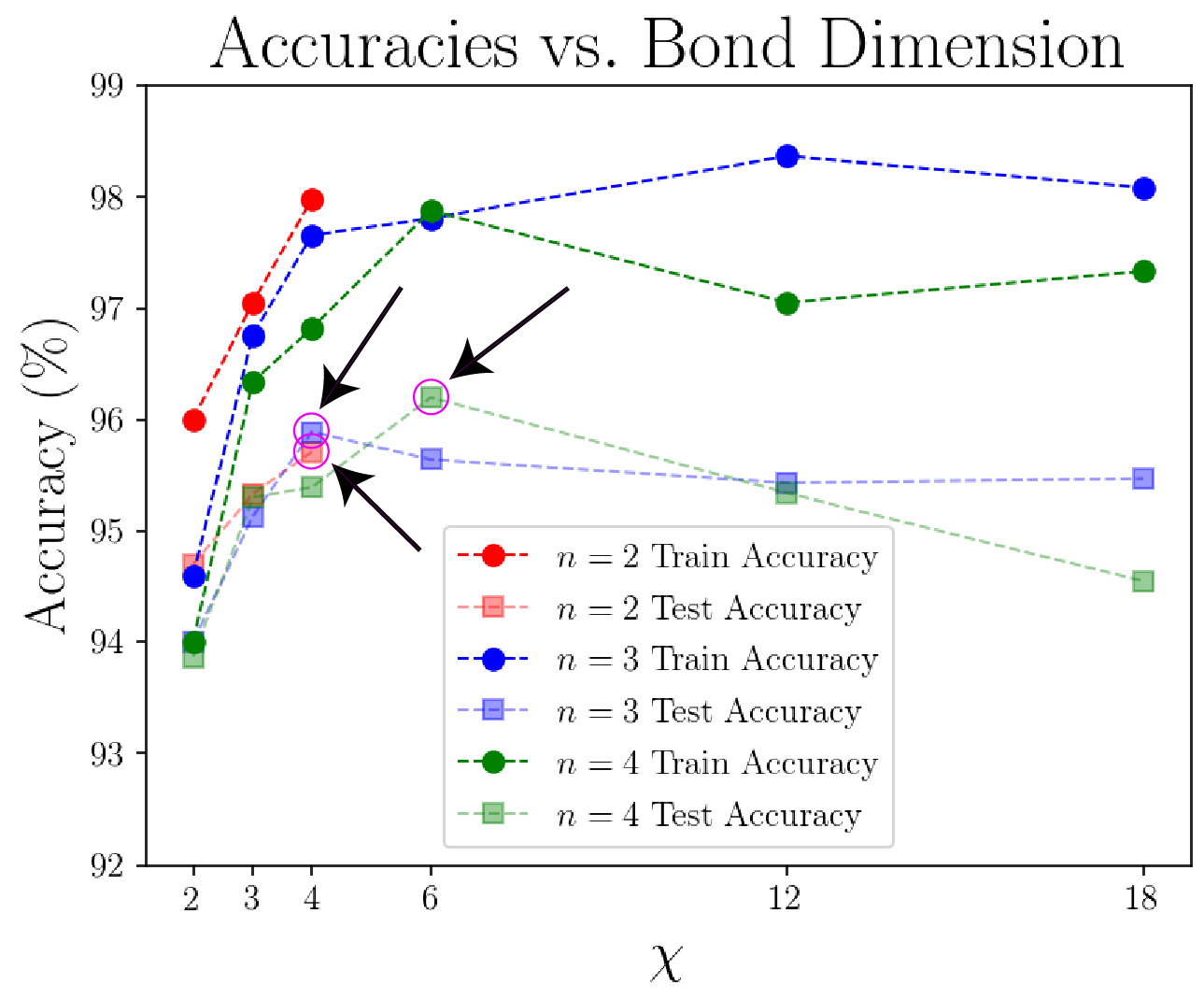}}
\caption{Training and test accuracies of snake block product states (SBPS) applied to the MNIST dataset of handwritten digits. The arrows point at the maximal test accuracy obtained for each size $n \times n$ of the blocks, for $n=2,3,4$. Suggestively, the maximal test accuracy is seen to increase monotonically with $n$.}
\label{fig:BPS_MPS_Acc}
\end{figure}

From this data, we see that the gap between training accuracy and test accuracy has closed significantly compared to the NNBPS model analysed above. This is due in part to a decrease in training accuracy, but also to an increase in test accuracy.  More specifically, starting with bond dimension $\chi=2$ both the train and test accuracy increase for small but increasing values of $\chi$. However, as the bond dimension grows further, the training accuracy generally continues to grow, while the test accuracy reaches a peak and then starts to decrease, signaling again over-fitting. Overall, however, SBPS is seen to perform better than NNBPS, in the sense that it generalizes better and achieves greater test accuracy.

We also report that using a redundant parameterization of the MPS tensor (e.g. a bond dimension larger than needed near the boundary of the MPS, such as a value larger than 2 for the bond dimension of the first or last MPS tensor) results, surprisingly, in an improved performance. We interpret this counter-intuitive result as indicating that there is clear room for improving test accuracies using alternative optimization schemes.

\section{Discussion} \label{sec:V}

In this work, we have conducted two different investigations that aimed to shed light into the role of entanglement in supervised image classification with tensor networks. In these approaches, each image is encoded as a vector in a vector space whose dimension is exponentially large in the number of pixels in an image. Then a tensor network is used to define a linear model in this massively large vector space, with a number of parameters that is only (roughly) proportional to the number of pixels in an image.

In the first investigation, we defined a sum state $\ket{\Sigma_{\ell}}$ as a superposition of all encoded images of class $\ell$ in the training set. We had imagined, incorrectly, that this state might be the one learned by e.g. the MPS in Ref. \cite{2016arXiv160505775M}. However, we found that the sum state is massively entangled. Approximating it by an MPS would require the bond dimension $\chi$ to be roughly equal to the number of images of class $\ell$ in the training set, which is about 6,000 images in MNIST, a number much greater than the largest MPS bond dimension $\chi = 120$ considered in Ref. \cite{2016arXiv160505775M}. We conclude that the tensor network model must be learning a state that is very different from the sum state $\ket{\Sigma_{\ell}}$.

In our second investigation, we defined block product states $\ket{\Psi^{\BPS}_{\ell}}$ that factorize into states $\ket{\psi_{\ell}^b}$ of blocks $b$ made of $n\times n$ qubits. By construction, these states only contain short-ranged entanglement -- entanglement within each $n\times n$ block of qubits. We then noticed that even $n=2$ leads to very large training accuracy, close to $100\%$, but that the models suffered from over-fitting, leading to poor test accuracy. We managed to partially alleviate over-fitting and improve generalization by considering different tensor network representations within each block. However, further work is still needed before these very simple, yet surprisingly expressive states are turned into competitive models for supervised image classification. We could not carry such investigation here due to time constraints, but we hope that our partial findings are already useful to other researchers in the field.

Entanglement plays a clear-cut role in the use of tensor networks for quantum many-body systems, where ground states of local Hamiltonians obey the so-called area law of entanglement entropy, that tensor networks can match. In contrast, much less is known about the role that entanglement plays in tensor networks for machine learning. However, in this work we have learned that, despite of the fact that entanglement is clearly useful -- notice that the training accuracy increased significantly in our block product states in going from $n=1$ (unentangled state) to $n=2$ (state entangled within blocks of $2\times 2$ qubits) -- large amounts of entanglement and long range may not be needed at all.

\textit{Acknowledgements}: J.M. and G.V. thank Cutter Coryell, Carlos Fuertes, Anna Golubeva, and Guy Gur-Ari for advice and thoughtful discussion. 

X, formerly known as Google[x], is part of the Alphabet family of companies, which includes Google, Verily, Waymo, and others (www.x.company).

\bibliography{References}
\bibliographystyle{apsrev4-1}

\appendix
\section{Schmidt spectrum and entanglement entropy} \label{sec:trick}

In this appendix we detail a method for calculating the Schmidt coefficients $\{\lambda_{\alpha}\}$ of the sum state $\ket{\Sigma_\ell}$ in Eq. \eqref{eq:Sigma_l}, from which we can easily also extract the entanglement entropy $S(A) = - \sum_{\alpha=1}^{\chi} (\lambda_{\alpha})^2 \log\left( (\lambda_{\alpha})^2 \right)$. More generally, we consider $N$ qubits in a state of the form 
\begin{equation}
    \ket{\Sigma}  =  \sum_{i=1}^{N_\Sigma} \ket{x^{(i)}},
\end{equation}
where the $N_\Sigma$ states $\{\ket{x^{(i)}}\}$ are product states (for instance, each product state $\ket{x^{(i)}}$ could be the result of applying a local feature map to an $N$-pixel image $x^{(i)}$, as discussed in Sec. \ref{sec:PrevWork}, although the specific origin of $\ket{x^{(i)}}$ is not relevant here). The manipulations below carry a computational cost that scales as $O(N_\Sigma^3)$, independently of the (potentially huge) dimension of the vector space of the $N$ qubits. Using this method one can compute the Schmidt coefficients for $N_\Sigma$ on the order of several thousands using a laptop.

\subsection{Schmidt decomposition}

Given an arbitrary partition of the $N$ qubits into two subsets $A$ and $B$, we can rewrite state $\ket{\Sigma}$ as
\begin{equation} \label{eq:product}
    \ket{\Sigma} = \sum_{i=1}^{N_\Sigma} \ket{x_A^{(i)}}\ket{x_B^{(i)}},
\end{equation}
where we use the fact that each product state $\ket{x^{(i)}}$ can be expressed as $\ket{x^{(i)}} = \ket{x_A^{(i)}}\ket{x_B^{(i)}}$. Alternatively, we can also rewrite $\ket{\Sigma}$ in its Schmidt decomposition,
\begin{equation} \label{eq:Schmidt2}
    \ket{\Sigma} = \sum_{\alpha=1}^{\chi} \lambda_{\alpha}\ket{\varphi^A_{\alpha}}\ket{\varphi^B_{\alpha}},
\end{equation}
where $\left\{\ket{\varphi^{A}_{\alpha}}\right\}$ and $\left\{ \ket{\varphi^{B}_{\alpha}} \right\}$ form orthonormal sets of vectors and the Schmidt rank $\chi$ is at most $N_\Sigma$.

\begin{figure}[H]
\center{\includegraphics[width=.49\textwidth]{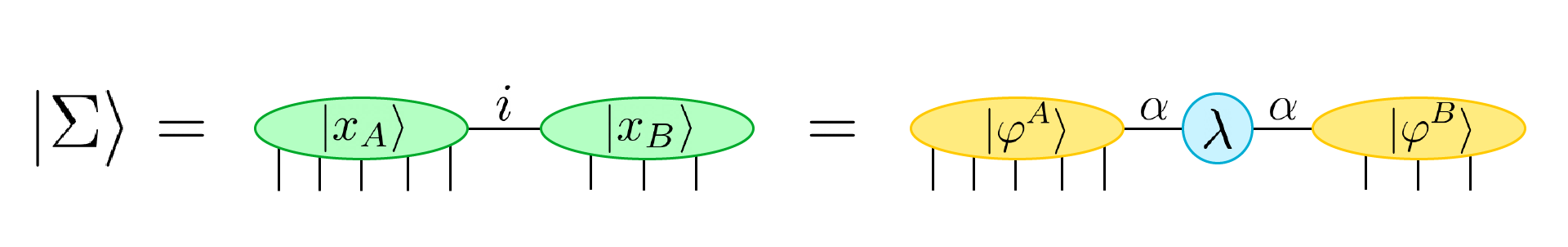}}
\caption{Schmidt decomposition of $|\Sigma\rangle$.}
\label{fig:Sigma1}
\end{figure}

To go from decomposition \eqref{eq:product} to decomposition \eqref{eq:Schmidt2} and extract the Schmidt coefficients $\{\lambda_{\alpha}\}$ we will proceed in two steps. First, we will map $\left\{\ket{x_A^{(i)}} \right\}$ into an intermediate orthonormal set  $\left\{ \ket{\psi^A_\gamma} \right\}$ of states on part $A$, 
\begin{equation}
 \ket{\psi^A_\gamma} = \sum_{i=1}^{N_\Sigma} \ket{x_{A}^{(i)}} (W_A)_{i \gamma},
\end{equation}
where $\gamma=1,\cdots, m$ for some $m\leq N_\Sigma$, by a change of basis given by an $N_\Sigma \times m$ matrix $W_A$ to be determined below.

\begin{figure}[H]
\center{\includegraphics[width=.49\textwidth]{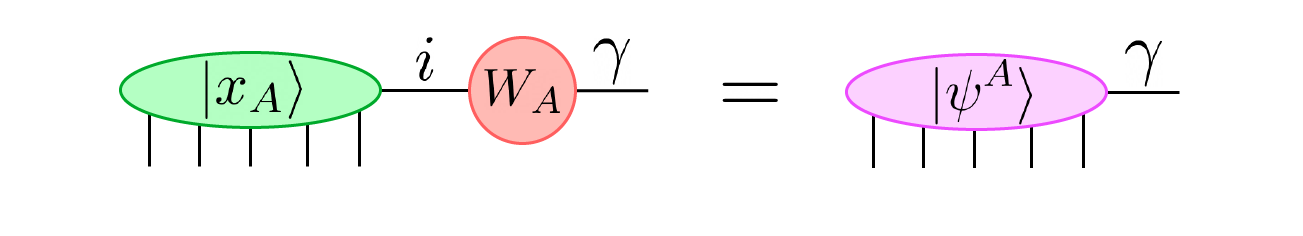}}
\caption{Construction of $|\psi^A\rangle$.}
\label{fig:psi_A}
\end{figure}

Similarly, we will map $\left\{ \ket{x_B^{(i)}} \right\}$ into an intermediate orthonormal set $\left\{ \ket{\psi^B_\gamma} \right\}$, 
\begin{equation}
 \ket{\psi^B_\gamma} = \sum_{i} (W_B^{T})_{\gamma i} \ket{x_{B}^{(i)}},    
\end{equation} 
by a change of basis given by some $m'\times N_\Sigma$ matrix $W_B^T$, where ${}^T$ denotes matrix transposition and $m' \leq N_\Sigma$.

\begin{figure}[H]
\center{\includegraphics[width=.49\textwidth]{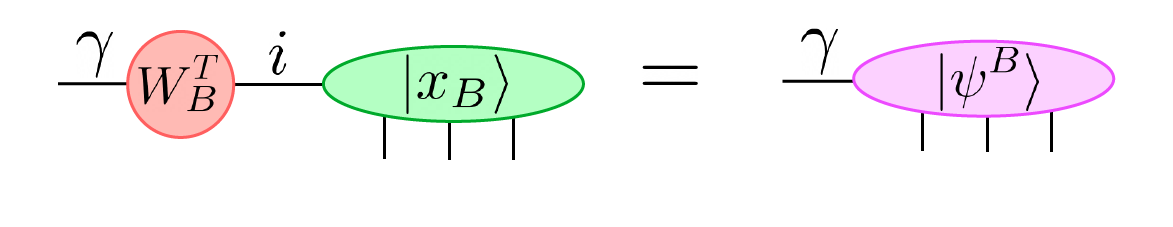}}
\caption{Construction of $|\psi^B\rangle$.}
\label{fig:psi_B}
\end{figure}

In terms of these orthonormal sets of vectors, state $\ket{\Sigma}$ reads
\begin{equation}
    \ket{\Sigma} = \sum_{\gamma=1}^{m}\sum_{\gamma'=1}^{m'} M_{\gamma \gamma'}  \ket{\psi^A_\gamma} \ket{\psi^B_{\gamma'}},
\end{equation}
with $M$ an $m\times m'$ matrix given by
\begin{equation} \label{eq:M}
    M = (W_A^{-1})(W_B^{-1})^{T}.
\end{equation}
Here $W_A^{-1}$ and $W_B^{-1}$ are (pseudo-)inverses of $W_A$ and $W_B$ such that
\begin{eqnarray}
\ket{x_{A}^{(i)}} &=& \sum_{\gamma} \ket{\psi^A_\gamma} (W_A^{-1})_{\gamma i},\\
\ket{x_{B}^{(i)}} &=& \sum_{\gamma} \ket{\psi^B_\gamma} \left((W_B^{-1})^{T}\right)_{i \gamma}.
\end{eqnarray}
Then, from the singular value decomposition of $M$, 
\begin{equation}
    M = V_A S V_B^{\dagger},
\end{equation}
we obtain the Schmidt values $\lambda_{\alpha}$ as the singular values of $M$ (given by the diagonal entries $S_{\alpha\alpha}$ of matrix S) whereas the Schmidt vectors read
\begin{eqnarray}
\ket{\varphi^{A}_{\alpha}} &=& \sum_{\gamma} \ket{\psi^{A}_{\gamma}}(V_{A})_{\gamma \alpha} \\
&=& \sum_{i} \ket{x^{(i)}_A} (W_AV_{A})_{i\alpha} \\
\ket{\varphi^{B}_{\alpha}} &=& \sum_{\gamma} (V_{B}^{\dagger})_{\alpha \gamma} \ket{\psi^{B}_{\gamma}}\\
&=& \sum_{i} ( V_{B}^{\dagger}W_B^{T})_{\alpha i} \ket{x^{(i)}_B}.
\end{eqnarray}

\subsection{Matrices $W_A$ and $W_B$}

In order to find matrix $W_A$ above we first build the Hermitian, positive semi-definite $N_\Sigma \times N_\Sigma$ matrix $X_A$ of scalar products
\begin{equation}
    (X_A)_{ij} \equiv \langle x^{(i)}_A | x^{(j)}_A \rangle.
\end{equation}
We then compute its eigenvalue decomposition
\begin{equation}
    X_A = U_A D_A U_A^{\dagger},
\end{equation}
where $U_A$ is an $N_\Sigma \times m$ isometric matrix (that is, $U_A^{\dagger} U_A = \mathbb{I}_{m}$) and $D_A$ is an $m \times m$ diagonal matrix with the $m$ strictly positive eigenvalues of $X_A$ in its diagonal entries ($(D_A)_{\gamma\gamma} > 0$). Notice that $U_A$ (and $D_A$) can be obtained from a regular eigenvalue decomposition of $X_A$ by simply ignoring the columns (respectively, columns and rows) corresponding to vanishing eigenvalues). Finally we set
\begin{eqnarray}
    W_A &\equiv& U_A ~D_A^{-1/2},\\
    W_A^{-1} &=& D_A^{1/2}~ U_A^{\dagger}. 
\end{eqnarray}
Notice that $W_{A}^{\dagger} X_A W_{A} = D_A^{-1/2} U_A^{\dagger}U_A D_A U_A^{\dagger} U_A D_A^{-1/2} = \mathbb{I}_{m}$, and that that $W_A W_A^{-1}$ is a rank-$m$ projector.
 
Similarly, we find the change of basis matrix $W_B$ above by building the Hermitian, positive semi-definite $N_\Sigma \times N_\Sigma$ matrix $X_B$ of scalar products
\begin{equation}
    (X_B)_{ij} \equiv \langle x^{(i)}_B | x^{(j)}_B \rangle,
\end{equation}
by computing its eigenvalue decomposition
\begin{equation}
    X_B = U_B D_B U_B^{\dagger},
\end{equation}
where $U_B$ is an $N_\Sigma \times m'$ isometric matrix with $m' \leq N_\Sigma$ and $D_B$ is an $m' \times m'$ diagonal matrix with strictly positive diagonal entries, and by then setting
\begin{equation}
    W_B \equiv U_B D_B^{-1/2},
\end{equation}
so that $W_{B}^{\dagger} X_B W_{B} = D_B^{-1/2} U_B^{\dagger}U_B D_B U_B^{\dagger} U_B D_B^{-1/2} = \mathbb{I}_{m'}$. Notice that $W_B^{-1} = D_B^{1/2} U_B^{\dagger}$, so that $W_B W_B^{-1}$ is a rank-$m'$ projector. 

Above we actually used the transposed matrices
\begin{eqnarray}
W_B^{T} &=& D_B^{-1/2} ~U_B^{T},\\
(W_B^{-1})^T &=& U_B^{*}~ D_B^{1/2},
\end{eqnarray}
where $*$ denotes complex conjugation and we used that for a unitary/isometric matrix $U$ we have $U^{\dagger} \equiv U^{*T} = U^{-1}$ and therefore $(U^{-1})^T = U^{*}$. 

\begin{figure}[H]
\center{\includegraphics[width=.49\textwidth]{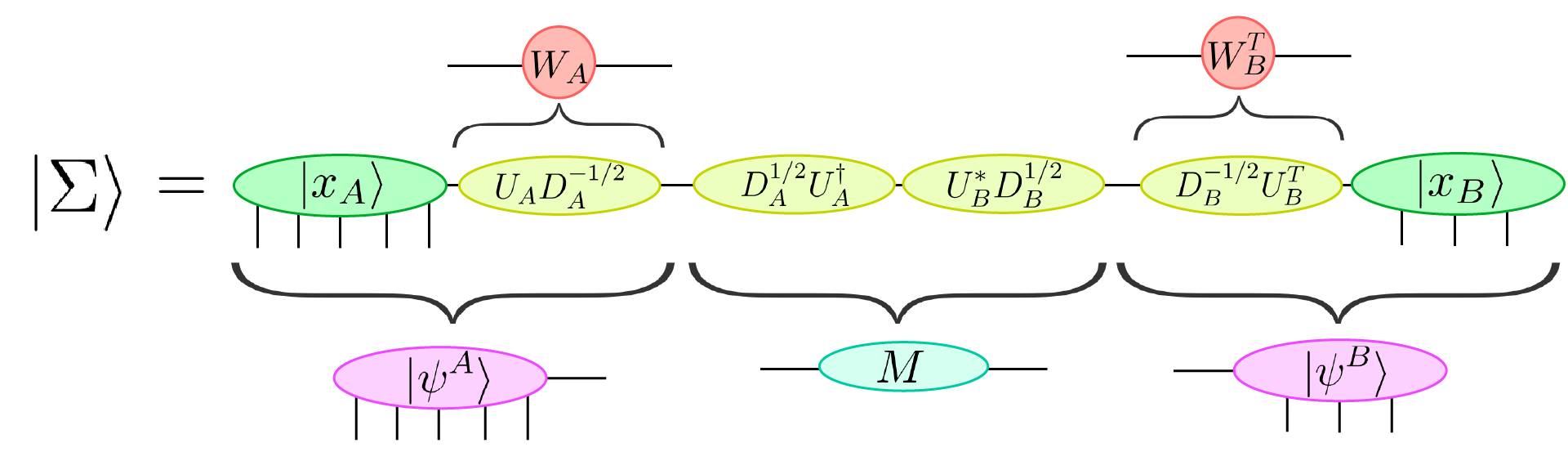}}
\caption{An equivalent expression for $|\Sigma \rangle$.}
\label{fig:s_ell_2}
\end{figure}

Finally, collecting all these terms together we can express the matrix $M$ in Eq. \eqref{eq:M} as
\begin{equation}
    M = (W_A^{-1})(W_B^{-1})^{T} = D_A^{1/2}~ U_A^{\dagger} ~U_B^{*} ~D_B^{1/2},
\end{equation}
whereas the Schmidt bases read
\begin{eqnarray}
\ket{\varphi^{A}_{\alpha}} &=& \sum_{i} \ket{x^{(i)}_A} (W_AV_{A})_{i\alpha} \\
&=& \sum_{i} \ket{x^{(i)}_A} (U_A~D_A^{-1/2}~V_{A})_{i\alpha}, \\
\ket{\varphi^{B}_{\alpha}}  &=& \sum_{i} ( V_{B}^{\dagger}W_B^{T})_{\alpha i} \ket{x^{(i)}_B} \\
&=& \sum_{i} (V_{B}^{\dagger} ~D_B^{-1/2}~U_B^{T})_{\alpha i} \ket{x^{(i)}_B}.
\end{eqnarray}
 
Importantly, we can build and diagonalize matrices $X_A$ and $X_B$, and build and singular value decompose matrix $M$ with a cost at most $O(N_\Sigma^3)$.

\begin{figure}[H]
\center{\includegraphics[width=.47\textwidth]{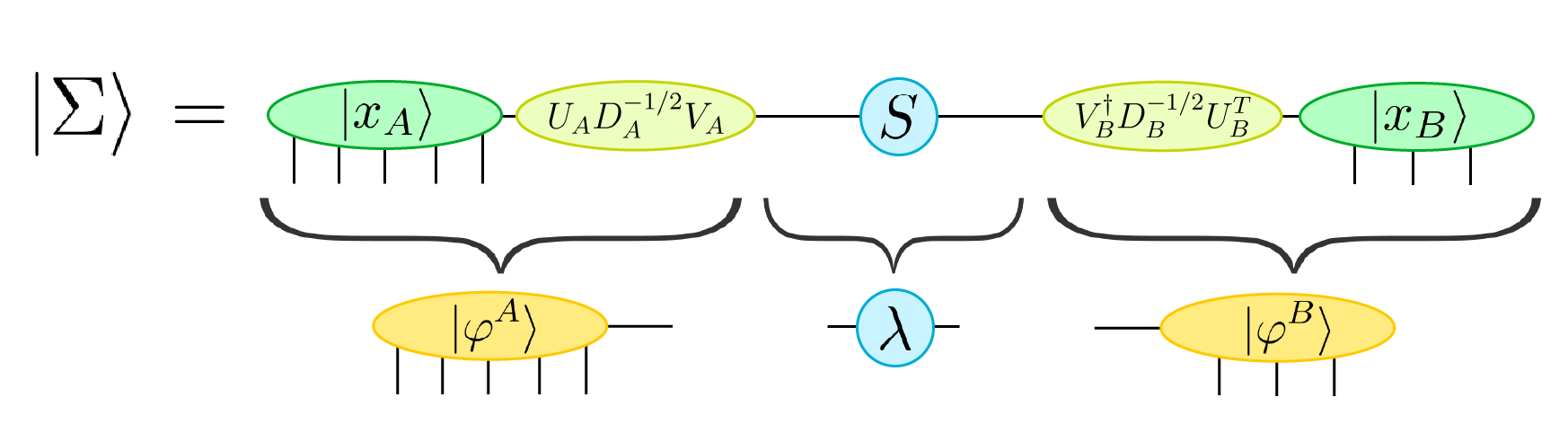}}
\caption{Another equivalent expression for $|\Sigma \rangle$. This expression elucidates how we can obtain the Schmidt coefficients $\lambda_\alpha$.}
\label{fig:s_ell_3}
\end{figure}

\end{document}